\documentclass[10pt, conference, compsocconf]{IEEEtran}
\IEEEoverridecommandlockouts
\usepackage{cite}
\usepackage{amsmath,amssymb,amsfonts}
\usepackage{algorithmic}
\usepackage{graphicx}
\usepackage{textcomp}
\usepackage{xcolor}
\def\BibTeX{{\rm B\kern-.05em{\sc i\kern-.025em b}\kern-.08em
    T\kern-.1667em\lower.7ex\hbox{E}\kern-.125emX}}
\begin{document}

\title{Global Income Inequality and Savings: A Data Science Perspective}


\author{\IEEEauthorblockN{Kiran Sharma}
\IEEEauthorblockA{School of Computational and Integrative Sciences\\
Jawaharlal Nehru University\\
New Delhi-110067, India. \\
kiransharma1187@gmail.com}
\and
\IEEEauthorblockN{Subhradeep Das}
\IEEEauthorblockA{School of Computational and Integrative Sciences\\
Jawaharlal Nehru University\\
New Delhi-110067, India. \\
subhradeep.das19@gmail.com}
\and
\IEEEauthorblockN{Anirban Chakraborti}
\IEEEauthorblockA{School of Computational and Integrative Sciences\\
Jawaharlal Nehru University\\
New Delhi-110067, India. \\
anirban@jnu.ac.in}
}

\maketitle

\begin{abstract}
A society or country with income equally distributed among its people is truly a fiction! The phenomena of socio-economic inequalities have been plaguing mankind from times immemorial. We are interested in gaining an insight about the co-evolution of the countries in the inequality space, from a data science perspective. For this purpose, we use the time series data for Gini indices of different countries, and construct the equal-time cross-correlation matrix. We then use this to construct a similarity matrix and generate a map with the countries as different points generated through a multi-dimensional scaling technique.  We also produce a similar map of different countries using the time series data for Gross Domestic Savings (\% of GDP). We also pose a different, yet significant, question: \textit{Can higher savings moderate the income inequality?} In this paper, we have tried to address this question through another data science technique -- linear regression, to seek an empirical linkage between the income inequality and savings, mainly for relatively small or closed economies. This question was inspired from an existing theoretical model proposed by Chakraborti-Chakrabarti (2000), based on the principle of kinetic theory of gases. We tested our model empirically using Gini index and Gross Domestic Savings, and observed that the model holds reasonably true for many economies of the world.

\end{abstract}


\section{Introduction}
\label{Introduction}

\begin{quote}
\textit{``We must work together to ensure the equitable distribution of wealth, opportunity, and power in our society.''}
\end{quote} 
\begin{flushright}
–-Nelson Mandela\\
-State of the Nation Address, Parliament, Cape Town, South Africa. \\
February 9, 1996
\end{flushright}

A society or country with income equally distributed among its people is truly a fiction! This socio-economic inequality has been a persistent fact and remains to be an elusive problem since time immemorial. Philosophers, religious leaders, social activists, academicians (including sociologists, economists and recently physicists), have passionately put their efforts in understanding the origin/cause and finding remedies to this multifaceted problem \cite{Rawls_2009,Scruton_1985,Sen_1983,Foucault_2003}. What has survived the tests of time is that the income inequality is a robust phenomenon, and in fact possesses certain statistical regularities \cite{Chakrabarti_2013}. Many studies have demonstrated that irrespective of the nature and size of the society, irrespective of the status of economy, irrespective of the time and geography of the country, we always observe empirically, a Maxwell-Boltzmann-Gibbs (or Gamma) distribution for the bulk, followed by a Pareto power-law tail in the income distribution \cite{Pareto_1897,Gibrat_1931}. In modern time, where everyone has his/her multidimensional perspective of looking at the problems/challenges, the question arises whether the tools of data science could be used to analyze the plethora of data available in order to shed some light over the one of the most fiercely debated topics in economics: Income inequality. The interdisciplinary field of data science uses scientific methods, processes, and algorithms to extract meaningful knowledge or insights from data in various forms.  In this paper, we are interested in gaining an insight about the co-evolution of the countries, both in the inequality space and savings space, from a data science perspective.

Further, we are interested in answering a different, yet significant, question -- \textit{ whether higher savings moderate the income inequality}, using a data science technique. This question took inspiration from a model proposed by Chakraborti and Chakrabarti (CC model) in 2000 \cite{Chakraborti_2000}, based on the statistical physics (kinetic theory) of ideal gases. The broad aim of statistical physics is to explain the physical properties of macroscopic systems, consisting of a large number of particles (of the order of Avogadro number $\sim 10^{23}$), in terms of the properties of microscopic constituents. Though it is extremely difficult to have a complete microscopic description of such a macroscopic system, because of the complexity of such systems, one can reliably estimate and relate the macroscopic observable quantities, which represent averages over microscopic properties \cite{Sethna_2006}. Indeed, the concepts and methods of statistical physics turned out to be extremely useful when applied to the understanding of diverse complex socio-economic systems \cite{Chakraborti_2015,Chakraborti_2011_a, Chakraborti_2011_b}, to the extent that all these studies have resulted in the interdisciplinary fields of ``econophysics'' \cite{Sinha_2010} and ``sociophysics'' \cite{Chakrabarti_2006, Sen_2013}. There are number of social phenomena which are now analysed and quite well-understood by physicists, e.g., a study of ethnic conflicts \cite{Sharma_2017_a}, a social phenomenon that is often rooted in socio-economic inequality, exhibits intriguing network properties and growth characteristics. Similarly, the dynamical nature of interactions of any granular economy, which is composed of a large number of cooperatively interacting agents at different levels (microscopic, mesoscopic and macroscopic) \cite{Sharma_2017_a}, has many features in common with the interacting complex systems \cite{Arthur_1999, Parisi_1999} that may be studied with the help of statistical physics. 

This paper is thus organized as follows: We analyse data on the Gini index \cite{Gini_1921}, which characterizes the income inequality quantitatively and the Gross Domestic Savings (\% of GDP) for European economies. The co-movements of the countries in savings and inequality space are captured using the multi-dimensional scaling (MDS) plots to understand similarity in evolution of the countries savings and inequality space. We then relate our findings to the saving propensity CC model of Kinetic Exchange Model (KEM) \cite{Chakraborti_2000,A.S.Chakrabarti_2009}, which had shown that for closed economies with positive savings propensity, the savings facilitate the reduction of inequality. Previously, Chakrabarti and Chakrabarti \cite{A.S.Chakrabarti_2010,Data_Worldbank_2017} analysed connections between the savings propensity parameter and measures of inequality like Gini index through numerical simulations. The present work provides an alternate empirical counterpart of the findings for selected economies. This hypothesis (if proven significant), along with the grouping in the inequality and savings spaces, would certainly play a crucial role in formulating better policies that are targeted towards reduction of inequality. 

\section{Data} 
\label{data}
As a measure of income inequality, we have considered the Gini indices for different countries in the world. This particular index has been considered, because it is scale independent, enabling us to directly compare two populations, regardless of their sizes.  As the measure of savings, we have studied  Gross Domestic Savings (\% of a GDP) data. All the data have been sourced from the World Bank database \cite{Data_Worldbank_2017}  and the Eurostat database \cite{Eurostat}. Due to availability of limited data, we had to analyze only a handful of countries, from where we have selected a group of countries (relatively small or closed economies) according to geography, size of economy and openness to trade. We would like to highlight the fact that anomalies (like missing or negative values) have been excluded from  GDS data. Negative savings indicate that countries spend more than what they earn as regular income and finance some of the expenditure, either by incurring debt or through gains arising from the sale of assets, or by running down savings that had been accumulated in the past. Since the CC model (explained in Section \ref{Model}) deals only with a positive savings propensity, we have neglected all the negative values of both the savings variable. 

\section{Empirical study}
\label{Empirical_Study}
Microscopic and macroscopic modeling help in imitating real socio-economic systems. There is a whole body of empirical evidence supporting the fact that a number of social phenomena are characterized by emergent behavior out of the interactions of many individual social components. Recently, the growing community of researchers have analysed large-scale social dynamics to uncover certain `universal patterns'. In this section we aim to study two aspects: (a) Co-movements of countries in inequality and savings spaces, and (b) Correlation between savings and inequality.

\subsection{Co-movements of countries in inequality and savings spaces }
\begin{figure*}
	\centering
	\includegraphics[width=0.9\textwidth]{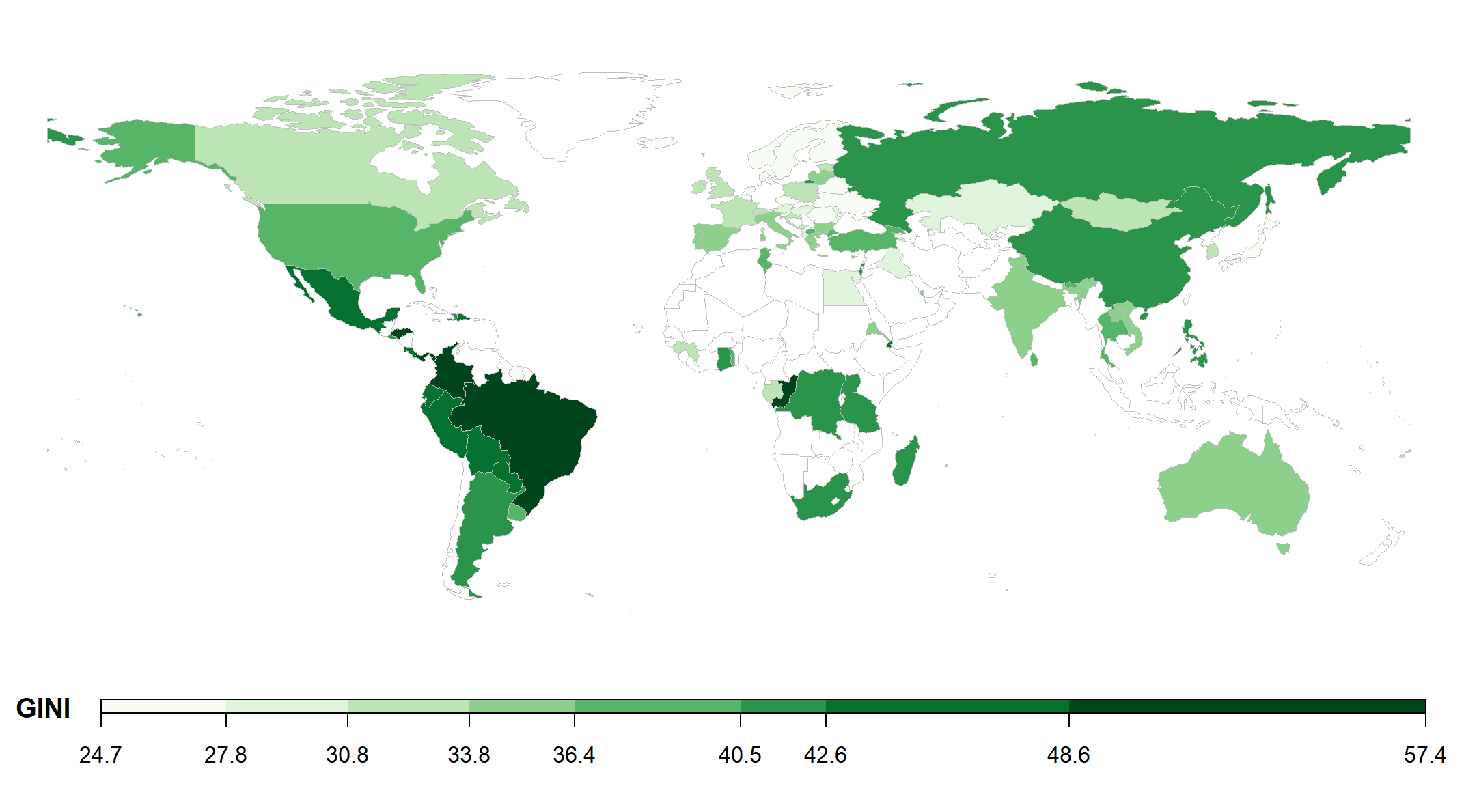}
\llap{\parbox[b]{6.4in}{(a)\\\rule{0ex}{3.5in}}}
	\vspace{1em}
	\includegraphics[width=0.9\textwidth]{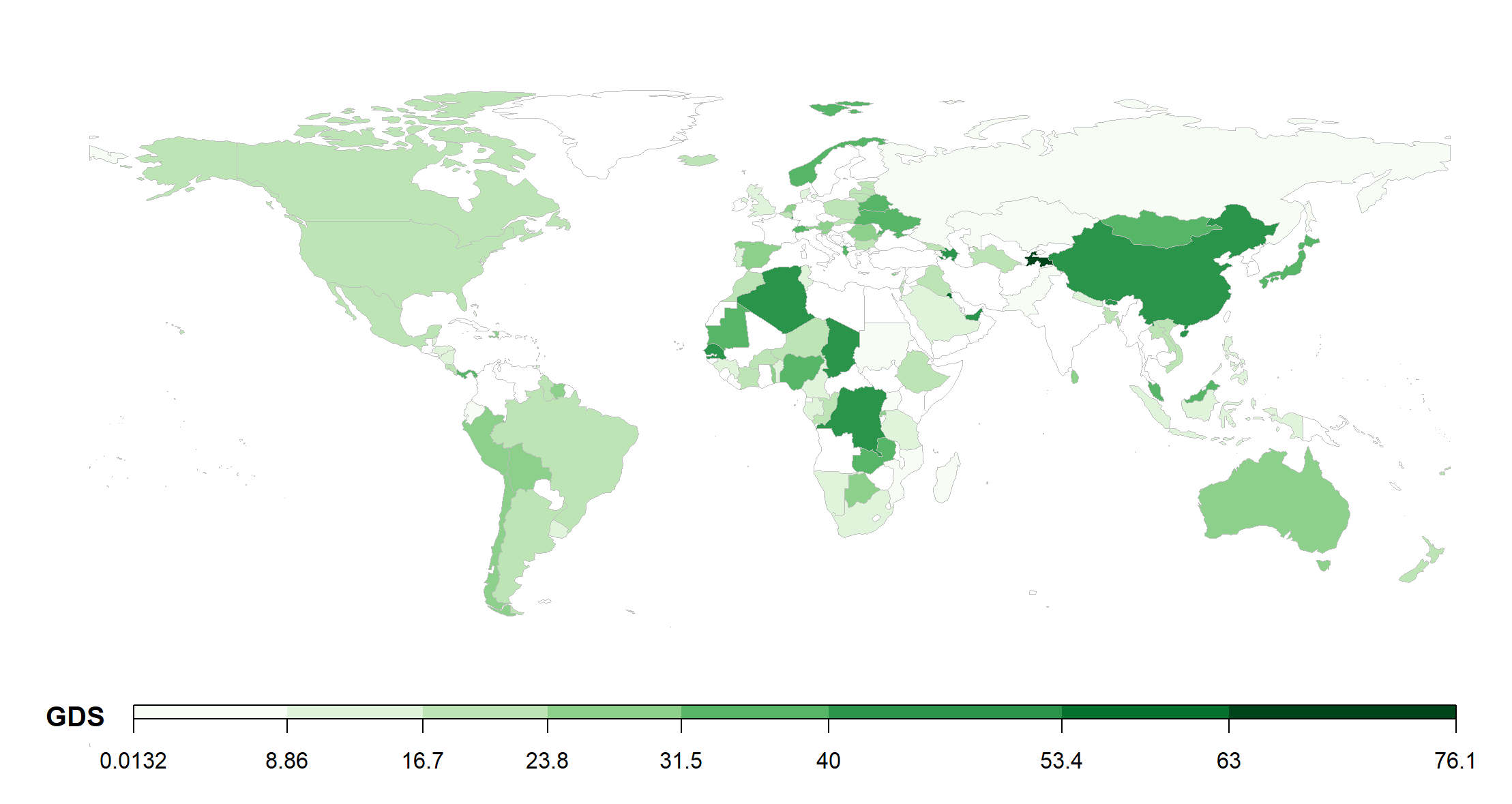}
\llap{\parbox[b]{6.4in}{(b)\\\rule{0ex}{3.5in}}}
\caption{The heat map representation for the regional distribution of (a) Gini Indices,  representing the countries' inequality and (b) Gross domestic savings (\% of GDP), on a world scale for the year 2012. Countries colored in light green represents a low value and darkgreen represents a high value, respectively. The world maps are generated by R-software. Note that the countries colored in \textbf{white represent missing data}. }
\label{Fig:world_map}
\end{figure*}

Various techniques have been proposed by scientists from varied fields to model and interpret inequality. The commonly used measures of socio-economic inequality are absolute, as, in terms of indices, for example: Gini, Theil, Pietra indices, which are represented by a single number. The alternative approach is relative in nature, i.e., using probability distributions of various quantities. Fig. \ref{Fig:world_map} visually represents the regional distribution of the Gini Indices  and gross domestic savings. Though a comparison can be charted out from the visual representation, the tool lacks the ability to incorporate the crucial element of time. Introducing the concept of time in studying the inequality and savings space would allow one to draw important insights from the pattern of co-evolution of economies.  One of the most efficient ways to model the evolution of such systems, is by using the toolbox of multidimensional scaling.

Using the time series data for Gini indices of different countries, we construct the equal-time cross-correlation matrix. We then use this to construct a similarity matrix and generate a map with the countries as different points with the help of the multi-dimensional scaling technique. This gives us an insight about the evolution of the countries in the inequality space. We also produce a similar map of different countries using the Gross domestic Savings (\% of GDP). All data analyses and numerical computing have been done using MATLAB programming. 


In order to study the equal-time cross-correlation matrix between $N$ countries, we compute the equal-time Pearson correlation coefficients $\rho_{ij}$ for each pair of countries $i$ and $j$. For this, we use the two time series data of length $T$, $c_i$ and $c_j$, for the countries $i$ and $j$, respectively. The  correlation coefficients $\rho_{ij}$ are mathematically defined as:
\begin{eqnarray}
\rho_{ij} = \frac{\langle c_i c_j \rangle - \langle c_i \rangle \langle c_j \rangle}{\sqrt{ [\langle c^{ 2}_i \rangle - \langle c_i \rangle^2][\langle c^{ 2}_j \rangle - \langle c_j \rangle^2]}}, 
\label{Eq:pearson_corr}
\end{eqnarray}
where $\langle...\rangle$ indicates an average over the length $T$. The correlation coefficients lies between $-1\leq \rho_{ij} \leq1$. Thus, we can create an $N \times N$ correlation matrix $C$ by collecting all values which are symmetric in nature, and gives us idea of which countries are moving together or in opposing directions.



To obtain ``similarities/dis-similarities", the following transformation
\begin{eqnarray}
 \zeta_{ij} = \sqrt{2 (1- \rho_{ij})},
 \label{Eq:distance}
\end{eqnarray}
is used, which satisfies all the propoerties of an ``ultrametric distance'' \cite{Rammal_1986} and $2 \geq \zeta_{ij} \geq 0$. Thus, we form an $N \times N$ similarities/dis-similarities matrix $S$ by  collecting all values of $\zeta_{ij}$ between countries $i$ and $j$.   



Multidimensional scaling is often used to display the structure of similarities/dis-similarities, given by Eq. \ref{Eq:distance}, as a geometrical map, where each country corresponds to a set of coordinates in a multidimensional space \cite{Borg_2005}. MDS arranges different countries in this space according to the similarities/dis-similarities between countries -- two similar countries are placed close to each other in the map, and two dissimilar countries are placed far apart. The minimum spanning tree (MST) is an unsupervised learning technique to (hierarchically) cluster similar objects, where the distances are given between all the objects. The MST is a graph which spans all the $N$ nodes by exactly $N-1$ edges, such that the sum of the distances of all the edges is a minimum \cite{Onnela_2003}.

\begin{figure}
	\centering
	\includegraphics[width=0.5\textwidth]{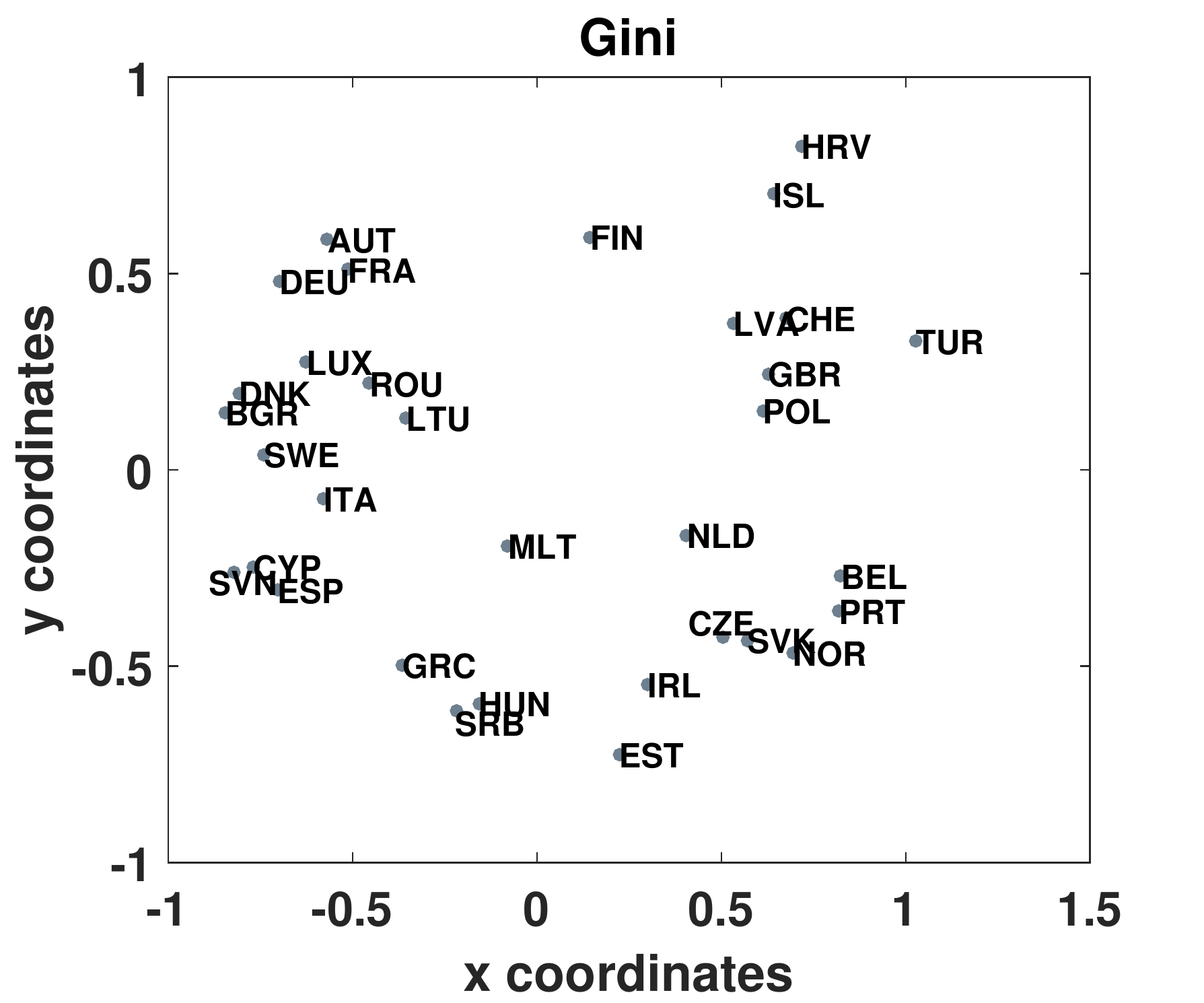}\llap{\parbox[b]{3.6in}{(a)\\\rule{0ex}{2.7in}}}
	\vspace{1em}
	\includegraphics[width=0.5\textwidth]{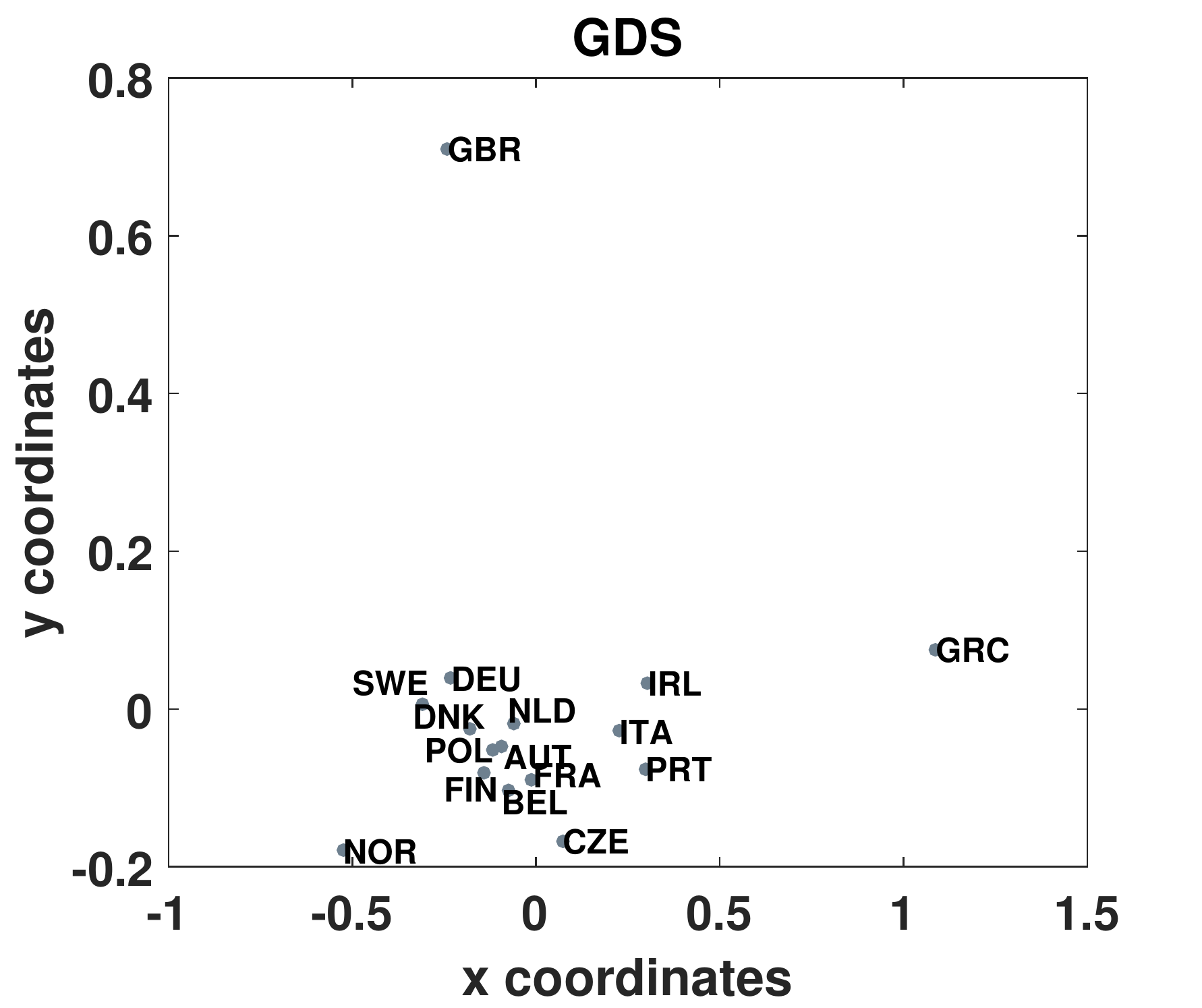}\llap{\parbox[b]{3.6in}{(b)\\\rule{0ex}{2.7in}}}
\caption{Co-evolution of countries in inequality space and saving space in the form of multidimensional scaling (MDS) map. (a) Using Gini indices for $33$ countries. (b) Using Gross Domestic Savings (as \% of GDP) for $20$ countries. The three letter country codes represent the countries, as listed in table \ref{tab:Country_List}.}
\label{Fig:MDS}
\end{figure}

\begin{figure}
	\centering
	\includegraphics[width=0.4\textwidth]{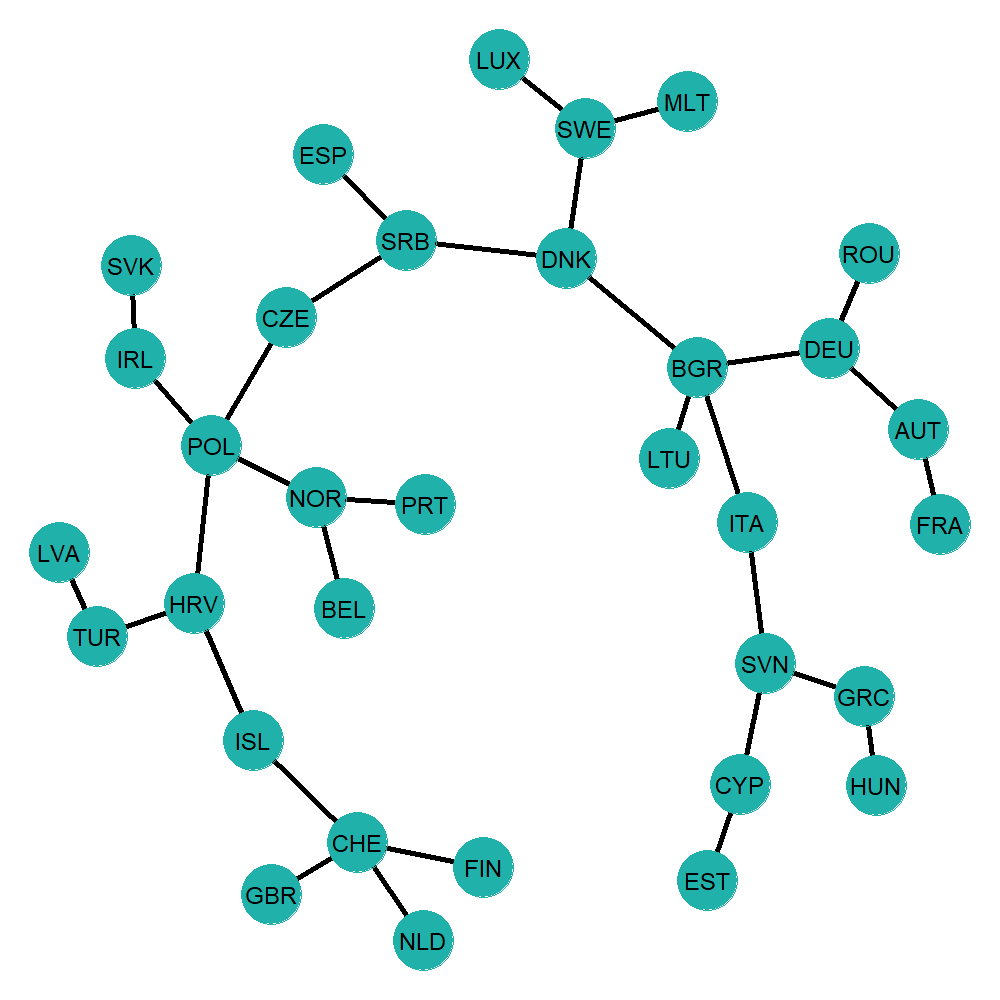}\llap{\parbox[b]{3in}{(a)\\\rule{0ex}{2.5in}}}
	\includegraphics[width=0.4\textwidth]{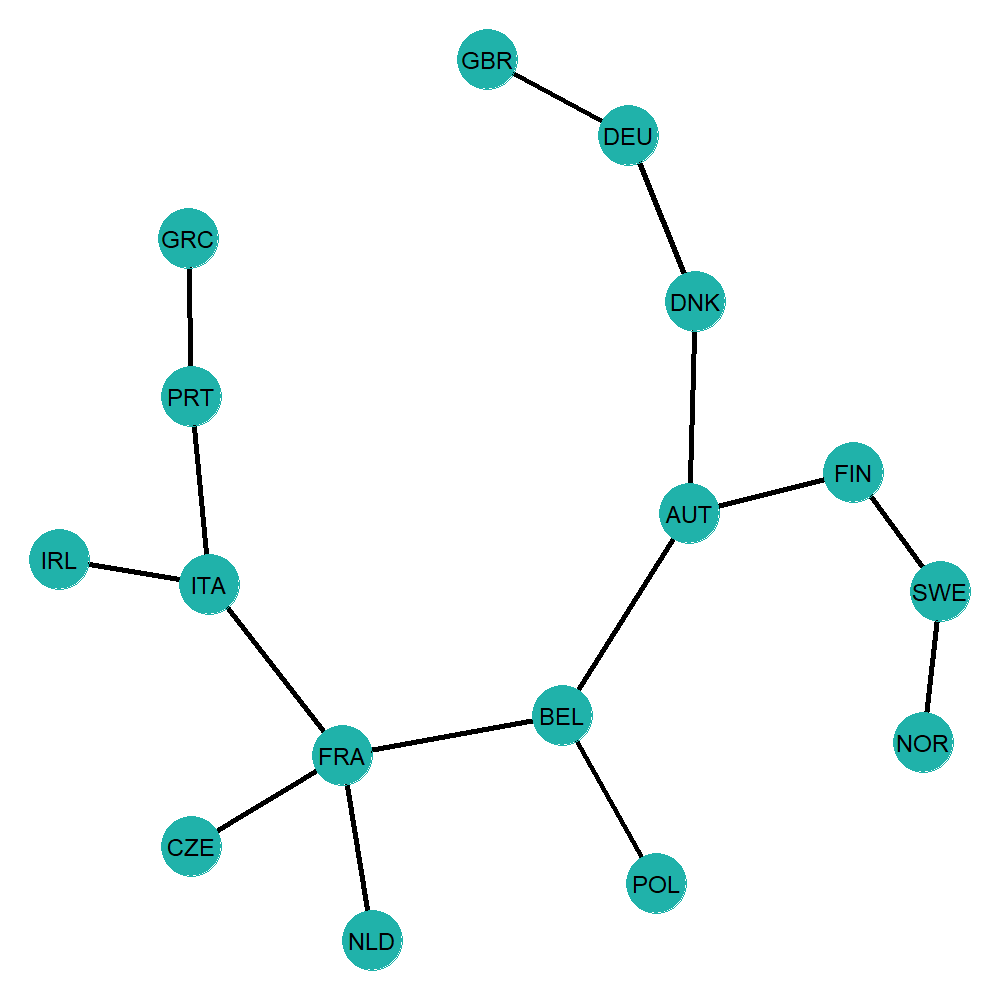}\llap{\parbox[b]{3in}{(b)\\\rule{0ex}{2.5in}}}
	\vspace{0cm}
\caption{Minimum spanning tree (MST -- hierarchical clustering) of countries in inequality space and saving space in the form of minimum spanning tree. (a) Using Gini indices for $33$ countries. (b) Using Gross Domestic Savings (as \% of GDP) for $20$ countries. The three letter country codes represent the countries, as listed in table \ref{tab:Country_List}.}
\label{Fig:MST}
\end{figure}

In order to capture the co-movement behavior of the countries visually, we have generated the MDS plots (see Fig. \ref{Fig:MDS}) and MST plots (see Fig. \ref{Fig:MST}) of countries, using the time series data for Gini indices and GDS.
The co-movements of countries in the savings and inequality spaces are very different, inferring that in income distribution and propensity to save evolve in different manners, even when countries belong to similar economic and political background. This can be attributed to the fact that other than economic variables, savings is also \textit{habit-driven}, and hence varies between societies, even if they are similar in nature in terms of economic and political background. On the hand, income distribution is mostly \textit{economy-driven} factors. From Fig. \ref{Fig:MDS} and Fig. \ref{Fig:MST}, it is clearly evident that there are certain pairs of countries which show similar co-movements in both the inequality and saving spaces, e.g., DNK-SWE and  FRA-AUT. On the other hand, ITA-PRT and FRA-BEL are very close in saving space but far away in inequality space. Understanding these co-movements would be important for policy making, and hence demands a thorough study.

\subsection{Correlation between savings and inequality}
\begin{figure}
	\centering
	\vspace{-0.6cm}
		\llap{\parbox[b]{0.2in}{(a)\\\rule{0ex}{-1in}}}
		\includegraphics[width=0.5\textwidth]{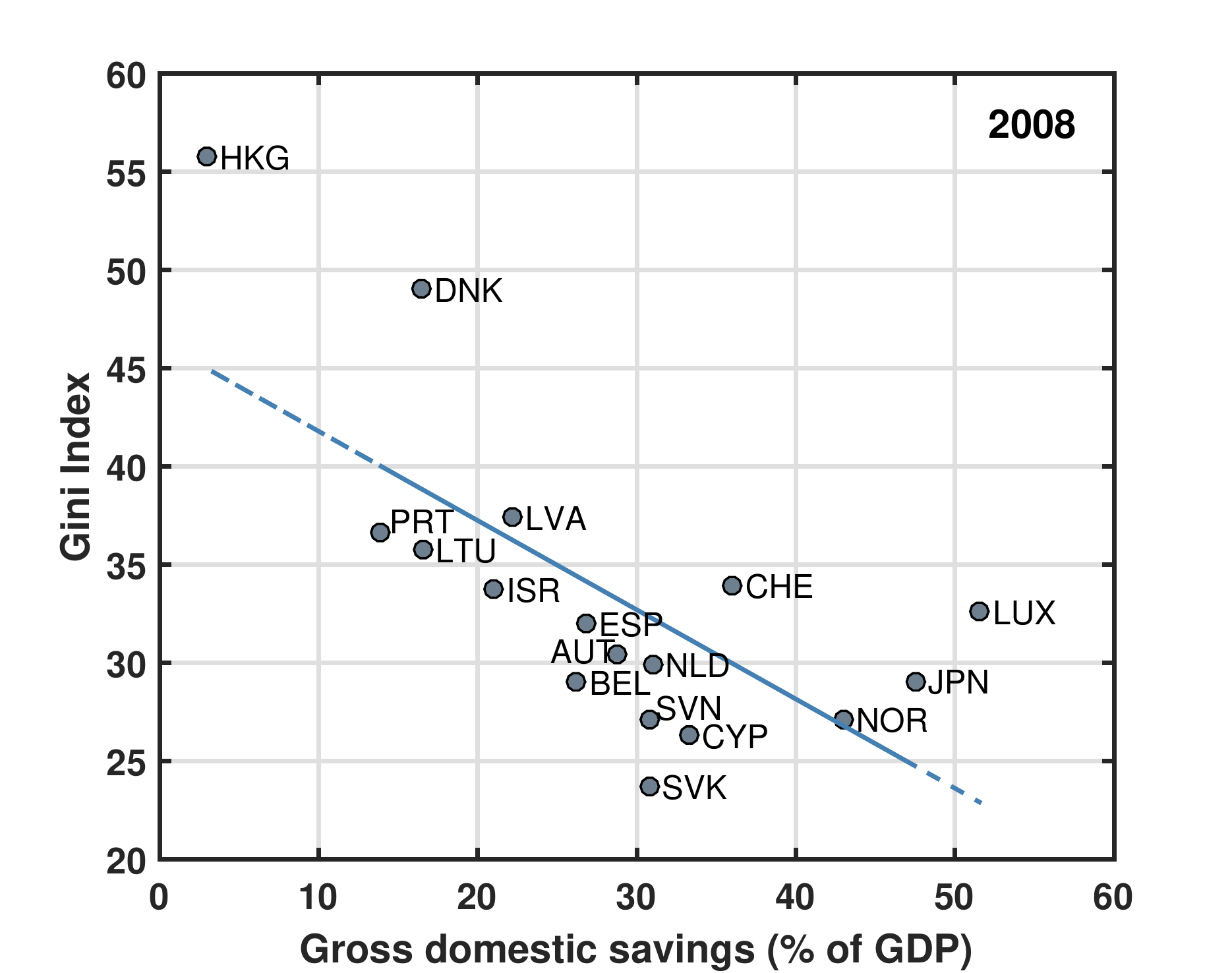}
		\vspace{-0.5cm}
	\llap{\parbox[b]{1.9in}{(b)\\\rule{0ex}{0in}}}
		\includegraphics[width=0.5\textwidth]{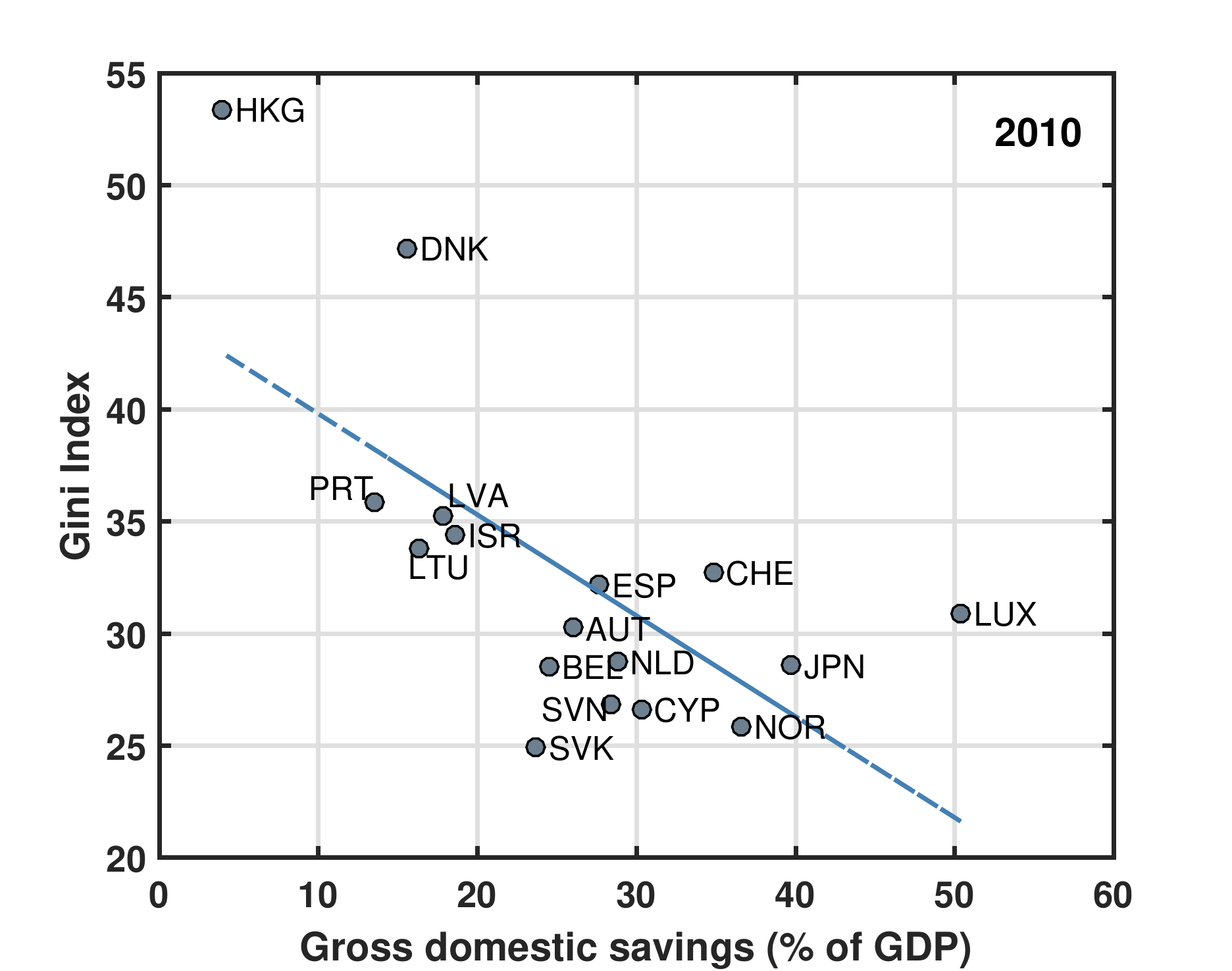}
		\vspace{-0.5cm}
\llap{\parbox[b]{1.9in}{(c)\\\rule{0ex}{0in}}}
		\includegraphics[width=0.5\textwidth]{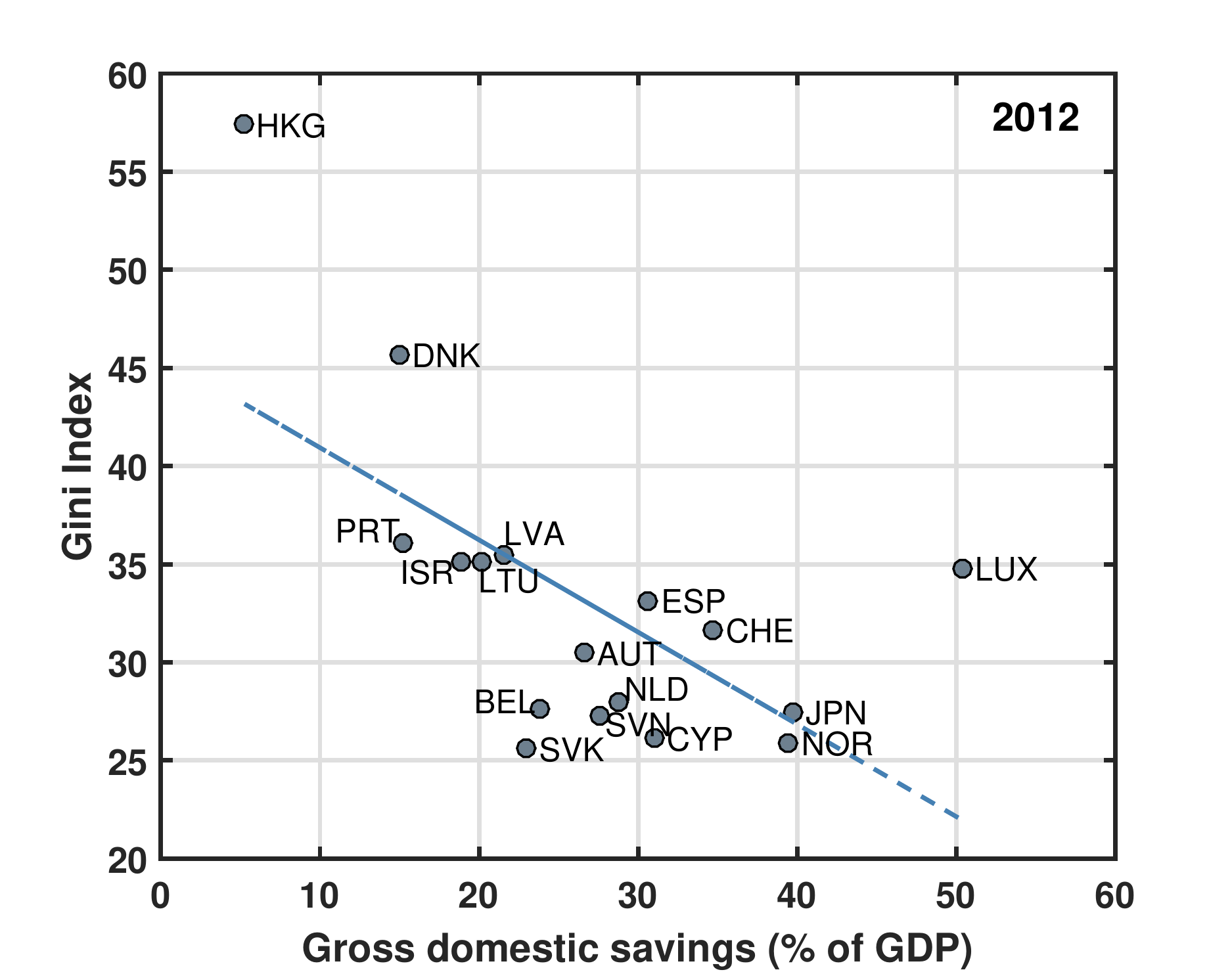}

\caption{Regression plots for GDS and Gini index. The three letter country codes represent the countries, as listed in table \ref{tab:Country_List}. (a) For the year 2008, with slope $  -0.45 \pm 0.12$ at p-value $0.002$. (b) For the year 2010, with slope $-0.45 \pm 0.13$ at p-value $0.003$. (c) For the year 2012, with slope $-0.47 \pm 0.15$ at p-value $0.007$.}

\label{Fig:GDS_GINI}
\end{figure}

Savings does play a very crucial role in a country's economy. The important relationship between savings and factors like economic growth have been long established by economists. Does it also play an important role in shaping the income distribution of a country, and thereby in the income inequality?
Here, we address this important question of whether the rate of savings can influence the income inequality, or not, using the Gini index and the GDS.  To verify this using empirical data, we have fitted a linear regression model with Gini index on GDS, using Ordinary Least Square (OLS) estimation, on the selected group of countries (listed in the Appendix) for the years
$2008$, $2010$ and $2012$. The results can be seen in Fig. \ref{Fig:GDS_GINI}.

Importantly, the slopes of both the regression lines indicate that GDS and Gini index are negatively associated (significant at less than $1\% $ level of significance). Hence, we can safely conclude that for an unit increase in gross savings of the economy there will be a drop in income inequality, as shown in the empirical data across different countries.

\begin{figure}
	\centering
		\includegraphics[width=0.5\textwidth]{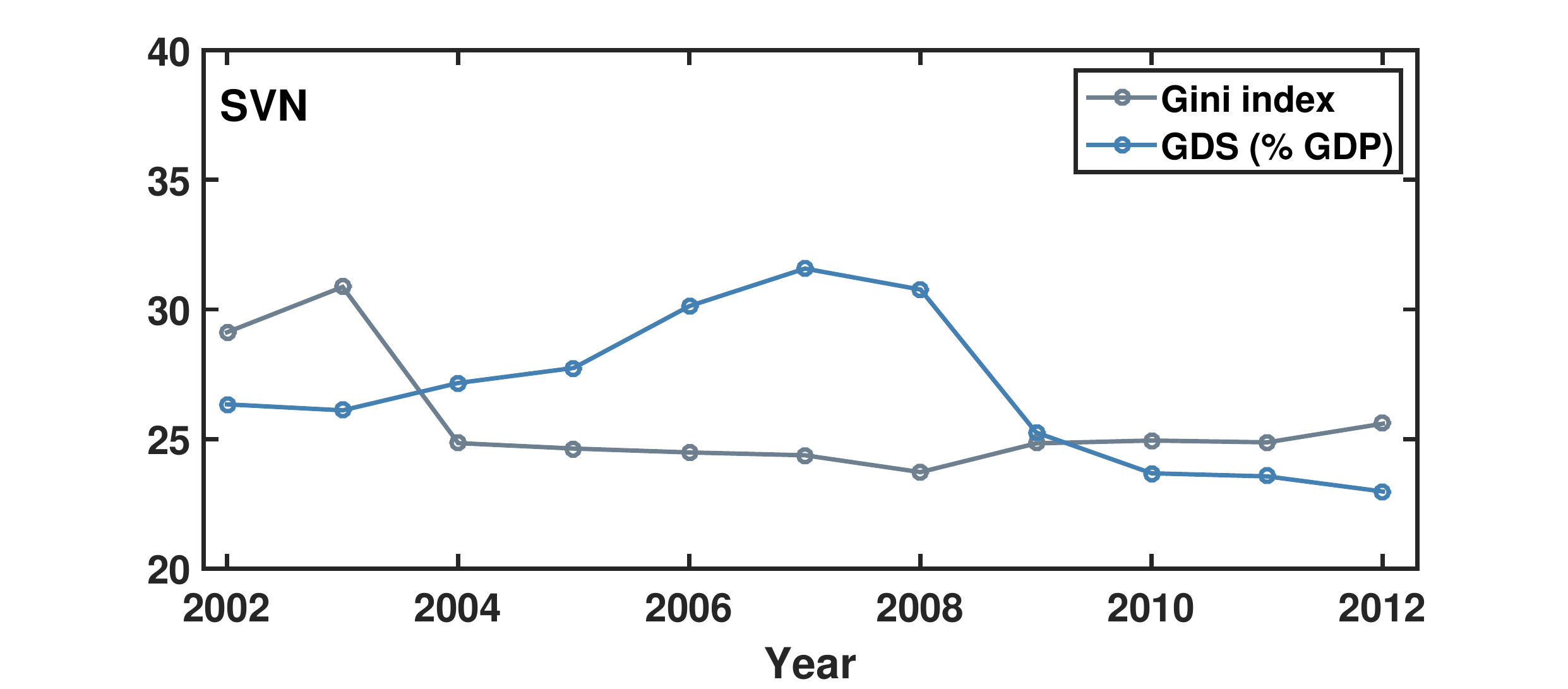}\llap{\parbox[b]{3.4in}{(a)\\\rule{0ex}{1.5in}}}
		\includegraphics[width=0.5\textwidth]{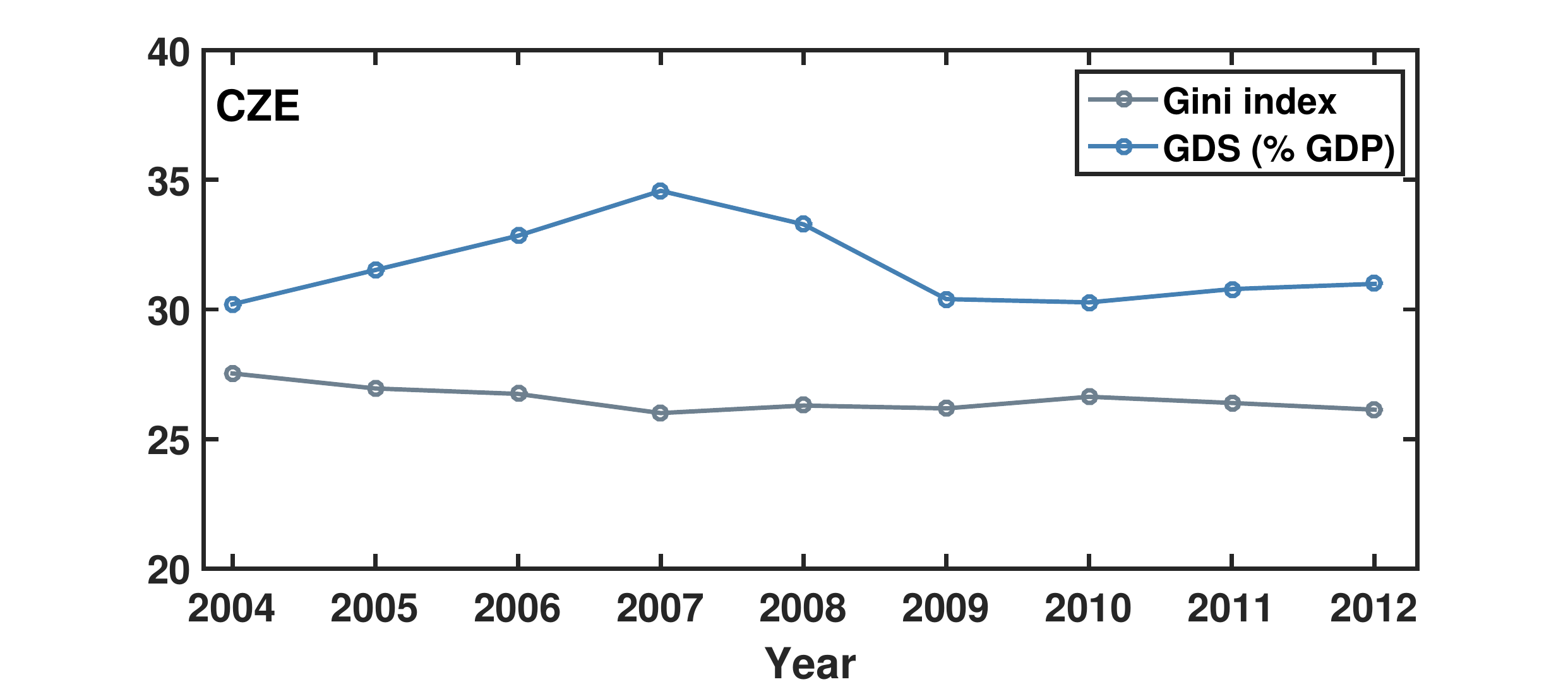}\llap{\parbox[b]{3.4in}{(b)\\\rule{0ex}{1.5in}}}
		
\caption{Plots of the time-series of Gini index and GDS (as \% of GDP) for the two countries: (a) Slovenia (SVN) having anti-correlation of $-0.27$, and (b) Czech Republic (CZE) having anti-correlation of $-0.42$.}
\label{Fig:SVN_CZE}
\end{figure}

We now focus in examining what kind of association exists between savings and income inequality within a country. Hence, we choose two time series of Gini indices and GDS for the countries, Slovenia (SVN) and Czech Republic (CZE), and evaluate the correlation between the two time series. We observe interesting results, as shown in Fig. \ref{Fig:SVN_CZE}, that the correlation coefficients for SVN $(-0.27)$  and CZE $(-0.42)$ are negative, implying that an increase in GDS is associated with a decline in Gini index. 

\section{Kinetic Exchange Model and numerical estimates}
\label{Model}
The question that does savings play an important role in shaping the income distribution of a country, and thereby in the income inequality, was inspired from a statistical physics model, based on the kinetic theory of gases \cite{Chakraborti_2000}. Thus, we now try to relate our empirical results to the Kinetic Exchange Models (KEMs), which are simple multi-agent models where the money exchanges (interactions) of autonomous agents (representing individuals, firms, business organizations, societies, countries, etc.), can be used to understand the collective behavior of the economic system. KEMs owe their popularity to the fact that they can capture many of the robust features of realistic income distributions using a minimal set of exchange rules.

Suppose we have $N$ agents in the closed economy who possess an initial amount of wealth $z_i (i =1,...,N)$. These agents transact at specific time intervals and an amount of wealth $\Delta z$ is exchanged between them. For any two agents $i$ and $j$, the transaction can be denoted using these simple equations: $z_i^\prime = z_i - \Delta z$; $z_j^\prime =  z_j + \Delta z$ by saying that the wealth is redistributed between the agents. For any point of time,  $z_i^\prime + z_j^\prime  = z_i + z_j$, i.e., the total wealth is conserved in the economy throughout all transactions. If the redistribution of wealth between two agents occurs pure randomly, then the basic model lead to an equilibrium Gibbs distribution. 
This distribution has often been deemed as an 
``unfair distribution" -- the majority of poor agents and a small minority of rich agents -- evident from the zero mode and the exponential tail. One possible explanation to such an unfair distribution is the wide inequality in skill distribution, with high skilled agents reaping greater benefit from the transactions than the low skilled agents. Another explanation can be the asymmetry in information that prevails in the economy. Agents possessing perfect information of the market will have an upper hand than agents possessing imperfect information. The former being negligible in number will tend to create a high inequality in wealth and income. Low or negligible savings can also be a reason for high inequality. If the agents in the economy have a high propensity to save, then they would be inclined towards retaining a portion of their income during each transaction that can lead to a lower inequality. We then consider CC model, where the effect of savings was introduced through a saving propensity $0 \leq \lambda < 1$, which represents the fraction of wealth that is saved – and not redistributed – during a transaction. The CC model with savings can be written as: 

\begin{eqnarray}
z_i^\prime &=& \lambda z_i + \epsilon(1-\lambda)(z_i+z_j) \nonumber\\
z_j^\prime &=& \lambda z_j + \bar{\epsilon}(1-\lambda)(z_i+z_j),
\end{eqnarray}

\noindent where $\bar{\epsilon} = 1- \epsilon $. This model leads to an equilibrium distribution (simulation data is well-fitted by the analytic distribution) of the form:

\begin{eqnarray}
  \phi_n(z) &=& a_n z^{n-1} \exp( - n z/\langle z \rangle) \ ,
  \nonumber \\
  a_n &=& \frac{1}{\Gamma(n)} \left( \frac{n}{\langle z \rangle} \right)^n\ ,
\label{NGibbs}
\end{eqnarray}
where the prefactor $a_n$ is fixed by the normalization condition $\int_{0}^{\infty} dx \phi_n(z) = 1$, $\Gamma(n)$ is the Gamma function and the parameter $n$ is defined as below:
\begin{equation}
  n(\lambda) = 1 + \frac{3 \lambda}{1 - \lambda} \ .
  \label{n}
\end{equation}
This particular form of $n(\lambda)$ was suggested by a mechanical analogy, 
discussed in Ref.~\cite{Patriarca_2004_b,Patriarca_2005,Patriarca_2006,Patriarca_2007,Chakraborti_2009,Patriarca_2013,Patriarca_2017}, between the closed economy model with $N$ agents and the dynamics of an ideal gas of $N$ interacting particles.
The distribution has a mode $z_m > 0$, which monotonically increases as a function of $\lambda$. 

\begin{figure}
	\centering
	\resizebox{0.5\textwidth}{!}{
			\includegraphics{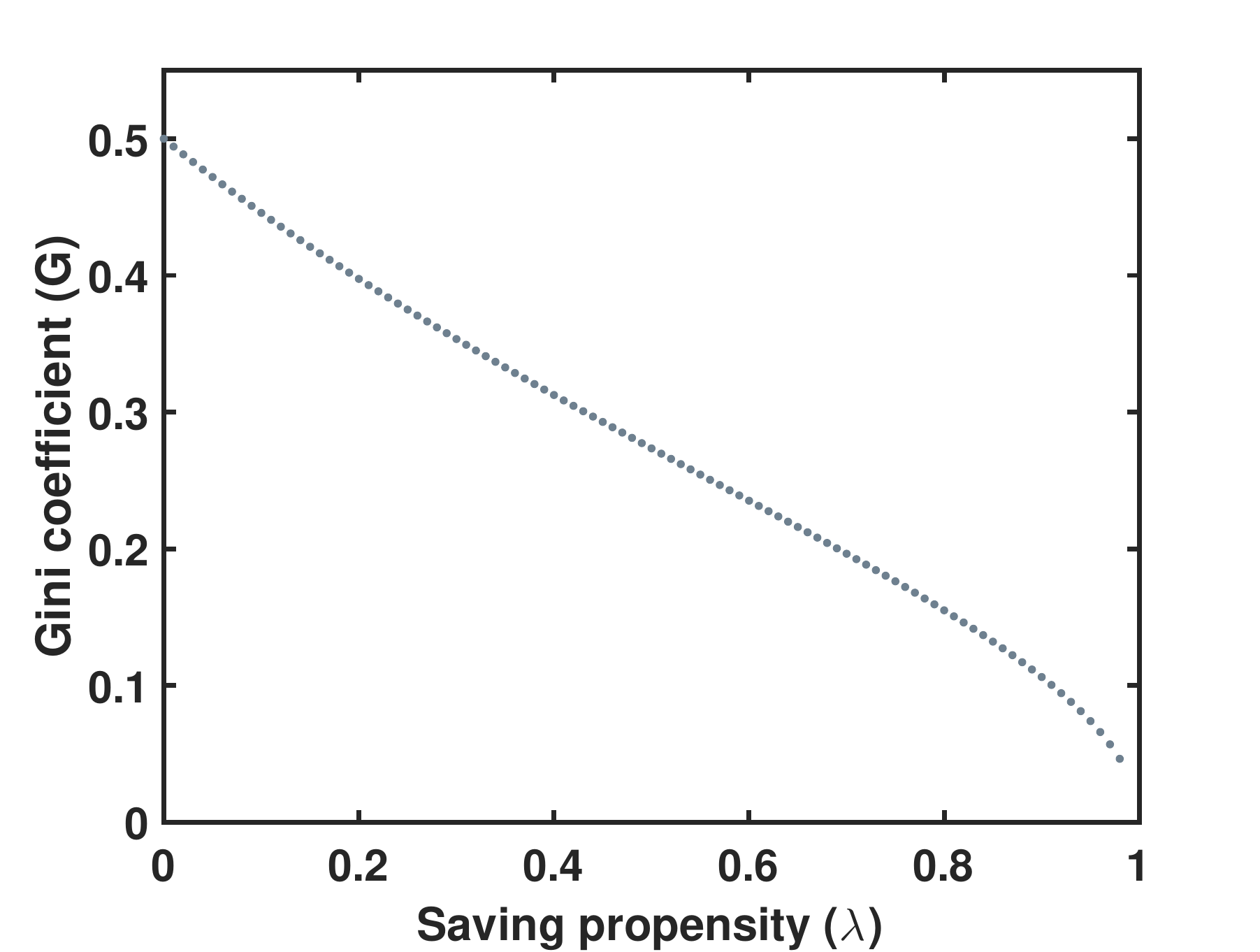}	
			}
\caption{The variation of the Gini coefficient (using numerical estimation from Eqs.~\ref{NGibbs}-\ref{Gini_Gamma}) with the saving propensity $\lambda$ (shown here for $0\leq \lambda \leq 0.99$). Note that as savings increases, the Gini coefficient decreases.}
\label{Fig:Gini_CCmodel}
\end{figure}

Interestingly, the CC model suggests that as $\lambda$ increases, the inequality in the distribution decreases. This can be captured by computing the Gini coefficient ($G$) for the cumulative distribution function $\Phi_n(y)=\int_{0}^{y} dx \phi_n(z)$, by the following relation \cite{Ghosh_2016,Chatterjee_2017}:
\begin{equation}
G(n)=1-{\frac {1}{\mu }}\int _{0}^{\infty }(1-\Phi_n(y))^{2}dy 
={\frac {1}{\mu }}\int _{0}^{\infty }\Phi_n(y)(1-\Phi_n(y)),
\label{Gini}
\end{equation}
which for the distribution given by Eq.~\ref{NGibbs}, takes the form:
\begin{equation}
G=\frac {\Gamma\left({\frac{2n+1}{2}}\right)}{n\,\Gamma(n){\sqrt {\pi }}}.
\label{Gini_Gamma}
\end{equation}
Fig. \ref{Fig:Gini_CCmodel} shows how the theoretical Gini coefficient varies with the saving propensity $\lambda$ using numerical estimation from Eqs.~\ref{NGibbs}-\ref{Gini_Gamma}.

Therefore, the CC model theoretically suggests that the rate of savings would be anti-correlated with income inequality (Gini index), which is what was observed empirically, in the previous section.

\section{Summary and Discussions}
In this paper, we have studied the global income inequality (Gini index) and savings (GDS) from a data science perspective. There are several economic variables and factors which lead to inequalities. However, inspired by a physics model we studied here the influence of savings in possibly reducing income inequality. The empirical data analyses for different countries suggested a close association (anti-correlation) between savings and inequality, but one should be careful in drawing further inferences or causal relations. More detailed studies are required to understand the mechanism of how savings actually reduces inequality. Here, we have studied countries mostly across Europe. It would certainly be useful to extend the studies where inequalities are severe, or the economies are different. For example, in developing countries, the lower income group has small or negligible savings propensity as compared to the high-income group. This further aggravates the problem of inequality as lower savings can lead to worsening of several key economic and social factors, e.g., the financial health.  As per our studies, if lower income group chooses to save more, the income gap may reduce after a considerable period of time, similar to what was suggested in Ref.~\cite{Vilalta_2012}. Though the economic model based on KEM is quite idealistic and limited to its assumptions, it surely captures the significance of savings in the context of inequality. 
If established to be true, as the empirical data suggests, then it will obviously have a widespread impact on policy formulation, especially for developing countries. Such countries experience high growth with a wide gap in wealth and income distribution. Among other policies that are targeted towards an egalitarian society, savings would play a crucial role. The findings from the MDS and MST plots raise a variety of questions that require further research and understanding. The co-movement of several group of countries, both in the inequality and savings spaces, might imply similar policy interventions to address the burgeoning problem of inequality, though a more detailed study is required to reach definitive conclusions. 
\section*{Acknowledgement}

AC and KS acknowledge the support by grant number BT/BI/03/004/2003(C) of Govt. of India, Ministry of Science and Technology, Department of Biotechnology, Bioinformatics division, University of Potential Excellence-II grant (Project ID-47) of JNU, New Delhi, and the DST-PURSE grant given to JNU by the Department of Science and Technology, Government of India. KS acknowledges the University Grants Commission (Ministry of Human Research Development, Govt. of India) for her senior research fellowship.

\section*{Appendix: List of countries and abbreviations}
The list of countries and their abbreviations are given in the Table~\ref{tab:Country_List}.
\begin{table}[h]
\centering
\caption{List of countries and abbreviations.}
\begin{tabular}{|l|l|l|l|l|l|}
\hline
S.No. & Code & Country        & S.No. & Code & Country  \\ \hline
1     & BEL  & Belgium        & 18 & MLT & Malta          \\ \hline
2     & BGR  & Bulgaria       & 19 & NLD & Netherlands    \\ \hline
3     & CZE  & Czech Republic & 20 & AUT & Austria        \\ \hline
4     & DNK  & Denmark        & 21 & POL & Poland         \\ \hline
5     & DEU  & Germany        & 22 & PRT & Portugal       \\ \hline
6     & EST  & Estonia        & 23 & ROU & Romania        \\ \hline
7     & IRL  & Ireland        & 24 & SVN & Slovenia       \\ \hline
8     & GRC  & Greece         & 25 & SVK & Slovakia       \\ \hline
9     & ESP  & Spain          & 26 & FIN & Finland        \\ \hline
10    & FRA  & France         & 27 & SWE & Sweden         \\ \hline
11    & HRV  & Croatia        & 28 & GBR & United Kingdom \\ \hline
12    & ITA  & Italy          & 29 & ISL & Iceland        \\ \hline
13    & CYP  & Cyprus         & 30 & NOR & Norway         \\ \hline
14    & LVA  & Latvia         & 31 & CHE & Switzerland    \\ \hline
15    & LTU  & Lithuania      & 32 & SRB & Serbia         \\ \hline
16    & LUX  & Luxembourg     & 33 & TUR & Turkey         \\ \hline
17 & HUN & Hungary       &  & & \\
\hline
\end{tabular}
\label{tab:Country_List}
\end{table}



\bibliographystyle{IEEEtran}
\bibliography{main}

\end{document}